\documentclass[12pt,preprint]{aastex}
\usepackage{graphicx}
\usepackage{rotating}

\shorttitle{Off-center Observers in Inhomogeneous Pressure Universes}
\shortauthors{A. Balcerzak et al.}

\received{}
\accepted{}
\journalid{}{}
\articleid{}{}

\begin{document}
\renewcommand{\theequation}{\thesection.\arabic{equation}}
\newcommand{\re}{\mathop{\mathrm{Re}}}
\newcommand{\be}{\begin{equation}}
\newcommand{\ee}{\end{equation}}
\newcommand{\bea}{\begin{eqnarray}}
\newcommand{\eea}{\end{eqnarray}}
\title{Off-center observers versus supernovae in inhomogeneous pressure universes}

\author{Adam Balcerzak, Mariusz P. D\c{a}browski, and Tomasz Denkiewicz}
\affil{\it Institute of Physics, University of Szczecin, Wielkopolska 15, 70-451 Szczecin, Poland}
\affil{\it Copernicus Center for Interdisciplinary Studies, S{\l }awkowska 17, 31-016 Krak\'ow, Poland}




\input epsf

\begin{abstract}

Exact luminosity distance and apparent magnitude formulas are applied to Union2 557 supernovae sample in order to constrain possible position of an observer outside of the center of symmetry in spherically symmetric inhomogeneous pressure Stephani universes which are complementary to inhomogeneous density Lema\^itre-Tolman-Bondi (LTB) void models. Two specific models are investigated. The first which allows a barotropic equation of state at the center of symmetry with no scale factor function being specified (model IIA), and the second which has no barotropic equation of state at the center, but has an explicit dust-like scale factor evolution (model IIB).

It is shown that even at $3\sigma$ CL, an off-center observer cannot be further than about 4.4 Gpc away from the center of symmetry which is comparable to the reported size of a void in LTB models with the most likely value of the distance from the center about 341 Mpc for model IIA  and 68 Mpc for model IIB. The off-center observer cannot be farther away from the center than about 577 Mpc for model IIB at $3\sigma$ CL. It is evaluated that the best-fit parameters which characterize inhomogeneity are: $\Omega_{inh} = 0.77$ (dimensionless - model IIA) and $\alpha = 7.31 \cdot 10^{-9}$ $(s/km)^{2/3} Mpc^{-4/3}$ (model IIB).

\end{abstract}



\maketitle

\section{Introduction}

Field theoretical discrepancy between observed and calculated values of the cosmological constant lead cosmologists to study non-Friedmannian models of the Universe which could explain the acceleration by the effect of inhomogeneity \citep{marrakolb07,UCETolman,center1,center2}. It was claimed that we lived in a spherically symmetric void of density described by the Lema\^itre-Tolman-Bondi (LTB) inhomogeneous dust spheres model \citep{L,T,B}. However, there is a variety of non-Friedmannian models \citep{BKC} which have the advantage that they are exact solutions of the Einstein field equations and may serve to solve the problem, too. Observational cosmology program of \citet{ellis} suggests that we should perhaps first start with model-independent observations of the past light cone, and then eventually make conclusions related to geometry of the Universe. Fundamentally, the homogeneity of the universe needs to be checked, and even if its large-scale structure is subject to the Copernican Principle after some averaging process \citep{buchert1,buchert2}, then one needs to prove that. In fact, non-Friedmannian models of the universe can fit observations very well and we need observational tools to differentiate between them and standard concordance models.

Assuming the spherical symmetry supported by cosmic microwave background (CMB) data can be the first step towards the task. In this context the inhomogeneous density $\varrho(t,r)$ (dust shells) LTB models are complementary to the inhomogeneous pressure $p(t,r)$ (gradient of pressure shells) Stephani models \citep{stephani,kras83,dabrowski93} since both of them are spherically symmetric and the only common part of them are Friedmann models which can be obtained in the limit of vanishing inhomogeneity. Because most of recent interest has concentrated onto the former models, then we would like to investigate such a complement of LTB models here. Accidentally, the Stephani universes were the first inhomogeneous models ever compared with observational data from supernovae \citep{dabrowski98} following the derivation of the redshift-magnitude relation for these models \citep{dabrowski95} which used the series expansion of \citet{KS} both for a centrally placed and an off-center observer. LTB models were first tested observationally by \citet{LTBtest1} and \citet{LTBtest2} and then followed more recently by \citet{biswas,MB2010,marra2011,valkenburg12}. It is worth mentioning that a general (non-spherically symmetric) Stephani model has no spacetime symmetries at all, though its three-dimensional hyperspaces of constant time are maximally symmetric like those of the Friedmann universe, and so it can be a good example of full inhomogeneity to study in future. A generalization of an LTB model which is fully spacetime inhomogeneous is the Szekeres model of which observational aspects have been studied recently by \citet{WH2012}. In this paper we restrict ourselves to the spherically symmetric Stephani models only.

Our paper is organized as follows. In Sec. \ref{Smodels} we briefly present some basic properties of inhomogeneous pressure Stephani models, also in comparison to the complementary LTB models. In Sec. \ref{off-center} we study an exact luminosity distance formula for an off-center observer in the Stephani universe. In Sec. \ref{models} we discuss some exact Stephani models useful for further discussion. Section \ref{data} contains the main result of the paper and deals with the constraints on the position of an observer who is away from the center of symmetry by the application of the Union2 supernovae data. In Sec. \ref{conclusions} we give conclusions.

\section{Inhomogeneous pressure Stephani universes}
\label{Smodels}

The spherically symmetric inhomogeneous pressure Stephani model is the only spherically symmetric solution of Einstein equations for a perfect-fluid energy-momentum tensor $T^{ab} = (\varrho + p) u^a u^b + p g^{ab}$ ($p$ is the pressure, $g^{ab}$ is the metric tensor, $u^a$ is the four-velocity vector) which is conformally flat (Weyl tensor vanishes) and embeddable in a five-dimensional flat space \citep{stephani}. It is complementary to an LTB spherically symmetric model in the sense that it has inhomogeneous pressure, while an LTB model has inhomogeneous density and the only common limit of both models is Friedmann. The metric of the spherically symmetric Stephani model reads as \citep{dabrowski93}
\bea
\label{STMET}
ds^2~=~-~\frac{a^2}{\dot{a}^2} \left[ \frac{ \left( \frac{V}{a} \right)^{\centerdot}}
  { \left( \frac{V}{a} \right)} \right]^2
c^2 dt^2~
+ \frac{a^2}{V^2} \left(dr^2~+~r^2 d\Omega^2
 \right),\nonumber \\
 &&
\eea
where
\be
\label{VSS}
  V(t,r)  =  1 + \frac{1}{4}k(t)r^2~,
\ee
and $(\ldots)^{\cdot}~\equiv~\partial/\partial t$. The function $V(t,r)$ with $k=0,\pm 1$ is of the same form for Friedmann models in isotropic coordinates (see Appendix A), $a(t)$ plays the role of a generalized scale factor, $k(t)$ has the meaning of a time-dependent ``curvature index,'' and $r$ is the radial coordinate. Kinematically, Stephani models are characterized by the nonvanishing expansion scalar $\Theta$ and  the acceleration vector $\dot{u}_a$. LTB models have non-zero expansion $\Theta$ and the shear tensor $\sigma_{ab}$.

\citet{dabrowski93} found two exact spherically symmetric Stephani models: model I which fulfills the condition $(V/a)^{\cdot \cdot}=0$, and model II which fulfills the condition $(k/a)^{\cdot}=0$. The metric for model II is simpler since the factor in front of $dt^2$ in the metric (\ref{STMET}) reduces just to $(-1/V^2)$. This  simplification will further be used in our paper to model the universe. Some models of type I have been investigated in a more detailed way by \citet{chris}.

The metric of the model II is given by \citep{dabrowski95,PRD13}
\be
\label{Steph1}
ds^2=-\frac{c^2dt^2}{V^2} + \frac{a^2(t)}{V^2}\left[ dr^2 + r^2(d\theta^2+\sin^2\theta d\varphi^2)\right]~.
\ee

\section{Luminosity distance for an off-center observer}
\label{off-center}

Making use of the standard relations for spherical coordinates
\be
\label{spher1}
x=r\sin\theta \cos\varphi ~, \, y=r\sin\theta \sin\varphi ~, \, z=r \cos\theta ~,
\ee
we transform the metric (\ref{Steph1}) into the Cartesian coordinate system as follows
\be
\label{Steph2}
ds^2=-\frac{1}{V^2}\left[-c^2dt^2 + a^2(dx^2+dy^2+dz^2)\right]~,
\ee
where $V(t,x,y,z) = 1 + (1/4)k(t)(x^2+y^2+z^2)$. Due to its fundamental property, the Stephani metric can be transformed to the form expressing its conformal flatness
\be
\label{Stephconf}
ds^2=\frac{a^2}{V^2}\left(-d\tau^2 + dx^2+dy^2+dz^2\right)~,
\ee
where we have used the conformal time
\be
\label{conftime}
d\tau=cdt/a(t)~.
\ee
Further, we apply another coordinate transformation of the form
\be
\label{x0}
x'=x-x_0~, \, y'=y-y_0~, \, z'=z-z_0~,
\ee
where $(x_0,y_0,z_0)$ is the position of an observer. The position of an observer at the center of symmetry is $(x_0,y_0,z_0) = (0,0,0)$. The metric (\ref{Stephconf}) in these new coordinates is simply
\be
\label{Stephprime}
ds^2=\frac{a^2}{V^2}\left[-d\tau^2 + dx'^2+dy'^2+dz'^2\right]~,
\ee
where
\be
V = 1 + \frac{1}{4} k(\tau) \left[ (x'+x_0)^2 + (y'+y_0)^2 + (z'+z_0)^2 \right]~.
\ee
We next transform the metric back to the spherical coordinate system, but now at the observer's position $(x_0,y_0,z_0)$ which is outside the center of symmetry, by the application of the spherical coordinates at this position
\be
\label{spher2}
x'=r'\sin\theta' \cos\varphi' ~, \, y'=r'\sin\theta' \sin\varphi' ~, \, z'=r' \cos\theta' ~,
\ee
which gives (\ref{Stephprime}) in the form
\be
ds^2=\frac{a^2}{V^2}\left[-d\tau^2 + dr'^2 + r'^2(d\theta'^2+\sin^2\theta' d\varphi'^2)\right]~.
\ee
In the new coordinate system $\{\tau, r', \theta', \varphi'\}$, all null geodesics that reach an observer at $r'=0$ fulfill the following conditions
\be
\label{geod}
d\tau=-dr'~, \, \theta'= const. ~, \, \varphi'=const~.
\ee
Suppose that an object (a supernova) is located at the distance $r'=\hat{r}'$ and has the coordinates $\theta'=\hat{\theta}'$ and $\varphi'=\hat{\varphi}'$ as seen by an observer placed at $(x_0,y_0,z_0)$. Then, the proper area of such an object is given by
\be
dS=\left[\frac{a^2(\tau)}{V^2(\tau,r')}\right]_{e} \hat{r}'^2\sin\hat{\theta}' d\hat{\theta}' d\hat{\varphi}'~,
\ee
where index ``e'' refers to an emitter of light (a supernova). Since the conformal factor $a^2/V^2$ preserves the angles measured the same both in flat and in curved spacetime, then the solid angle spanned by an object as seen by an observer is given by
\be
d\Omega= \sin\hat{\theta}' d\hat{\theta}' d\hat{\varphi}'~.
\ee
The area distance $d_A=\sqrt{dS/d\Omega}$ is
\be
\label{da1}
d_A=\left[\frac{a}{V}\right]_{e}\hat{r}'~.
\ee
The redshift in the Stephani universe (\ref{Steph1}) reads as \citep{dabrowski95,PRD13}:
\be
\label{rs}
1+z=\frac{a_0}{a_{e}}\frac{V_{e}}{V_0}~~,
\ee
where index ``0'' refers to the present. Due to the \citet{etherington} reciprocity theorem, we relate the luminosity distance $d_L$ with the area distance $d_A$ as
\be
d_L=(1+z)^2 d_A~~,
\ee
and so finally the luminosity distance is
\be
\label{dl}
d_L=\frac{a_0(1+z)\hat{r}'}{1+\frac{\beta}{4}a_0r_0^2}~.
\ee
Further, we will assume that $r_0$ indicates the position of an observer in the coordinate system $\{t,r,\theta,\varphi\}$ of the metric (\ref{Steph1}). Since the observational data is given in terms of the apparent magnitude, then we apply the standard relation
\be
\mu(z)= 5 \log(d_L)+25,
\ee

The same formula (\ref{dl}) for the luminosity distance can alternatively be obtained by using the area distance definition of \citet{ellis} which reads as
\be
\label{ellis}
d_A^4 \sin^2 \gamma = \tilde{g}_{\gamma\gamma} \tilde{g}_{\xi\xi}-   \tilde{g}^2_{\gamma\xi}~,
\ee
where $\tilde{g}_{\mu\nu}$ is the metric expressed in an observer's frame, i.e. a frame which is centered on the observer in which the angular part of the metric is given by
\be
\Omega^2=d\gamma^2+\sin^2\gamma d\xi^2~.
\ee
Here the angles $\gamma$ and $\xi$ correspond to the polar and azimuthal angles in this frame. We notice that an observer frame on which the formula (\ref{ellis}) relies on, precisely coincides with the frame described by the coordinates $\{\tau,r',\theta',\varphi'\}$ which were introduced in (\ref{spher2}), provided that we make the identifications: $\gamma\equiv\theta'$ and $\xi\equiv\varphi'$. With such identifications we immediately obtain that
\bea
\tilde{g}_{\gamma\gamma}&=&\tilde{g}_{\theta'\theta'}=\frac{a^2}{V^2}r'^2~, \\
\tilde{g}_{\xi\xi}&=&\tilde{g}_{\varphi'\varphi'}=\frac{a^2}{V^2}r'^2\sin^2\theta'~, \\
\tilde{g}^2_{\gamma\xi}&=& \tilde{g}^2_{\theta'\varphi'}~=0.
\eea
Finally, the application of (\ref{ellis}) gives
\be
d_A=\frac{a_e}{V_e}\hat{r}'~,
\ee
which coincides with (\ref{da1}).

\section{The models}
\label{models}

\subsection{Model IIA}

A subclass of model II with $k(t) = \beta a(t)$ ($\beta$ = const, with the unit $[\beta]$ = Mpc$^{-1}$) was proposed by \citet{stelmach01} and it was assumed that at the center of symmetry the standard barotropic equation of state $p(t) = w \varrho(t)$
was fulfilled. This assumption means that
\be
\label{SJmod}
 \frac{8\pi G}{3c^2} \varrho(t) = \frac{A^2}{a^{3(w+1)}(t)} \hspace{0.3cm} (A = {\rm const.})
\ee
and allows one to write a generalized Friedmann equation as
\be
\label{SJmod2}
 \frac{1}{c^2} \left(\frac{\dot{a}(t)}{a(t)} \right)^2 = \frac{A^2}{a^{3(w+1)}(t)} - \frac{\beta}{a(t)}
\ee
with the equation of state
\be
\label{weff}
p(t) = \left[w + \frac{\beta}{4} (w+1) a(t) r^2 \right] \varrho(t) = w_{e} \varrho(t)~~.
\ee
From (\ref{SJmod2}) and (\ref{weff}) we can see that the standard dust-filled ($w=0$) Friedmann universe solution $a(t) \propto t^{2/3}$ is possible outside the center of symmetry only in the limit $\beta \to 0$, i.e. when no inhomogeneity is present. In the next subsection we will present the solution with such a form of the scale factor which admits the inhomogeneity, but no barotropic equation of state at the center.

Similarly as in the Friedmann models, one can define the critical density as $\varrho_{cr}(t) = (3/8\pi G)[\dot{a}(t)/a(t)]^2~,$
and the density parameter $\Omega(t) = \varrho(t)/\varrho_{cr}(t)$. After taking $t = t_0$, we have from (\ref{SJmod})  that
\be
\label{defOM}
1 = \frac{A^2}{H_0^2 a^{3(w+1)}(t_0)} - \frac{\beta}{H_0^2 a_0} \equiv \Omega_0 + \Omega_{inh}~~,
\ee
and so
\be
\label{beta}
\beta = - a_0 H_0^2 \Omega_{inh} < 0~~,
\ee
where $H=\dot{a}/a$ is the Hubble parameter, the dimensionless parameter $\Omega_0$ stands for the barotropic matter content, while $\Omega_{inh}$ stands for the inhomogeneity density. A generalized Friedmann equation can be written as
\be
\label{friedeq}
\frac{H^2(t)}{H_0^2}=\Omega_0 a^{-3(w+1)}+\frac{\Omega_{inh}}{a}~,
\ee
where the form of the function $a(t)$ is not specified. Using (\ref{geod}), (\ref{friedeq}), and the definition of the conformal time, we find that
\be
\label{dist}
\hat{r}'=\hat{r}'(a)=\frac{1}{H_0}\int_{a_e}^1\frac{dx}{\sqrt{(1-\Omega_{inh})x^{1-3w}+\Omega_{inh}x^3}}~,
\ee
where $a_e$ is the value of the scale factor at the moment of an emission of the light ray. For the model (\ref{friedeq}), the redshift (\ref{rs}) reads as
\be
\label{rs2}
1+z=\frac{a_0 (4- a_e H_0^2 \Omega_{inh}r_e^2)}{a_e(4-a_0H_0^2 \Omega_{inh}r_0^2)}~,
\ee
with
\bea
\label{re}
r_e^2 &=& (r_0 \sin\theta_0 \cos\varphi_0+\hat{r}'(a)\sin\hat{\theta}' \cos\hat{\varphi}')^2  \nonumber \\
& +& (r_0 \sin\theta_0 \sin\varphi_0+\hat{r}'(a)\sin\hat{\theta}' \sin\hat{\varphi}')^2  \nonumber \\
& +& (r_0 \cos\theta_0 +\hat{r}'(a)\cos\hat{\theta}')^2
\eea
where $r=r_0$, $\theta=\theta_0$ and $\varphi=\varphi_0$ indicate the position of an observer in the very first coordinates of the metric (\ref{Steph1}), while $\hat{\theta}'$ and $\hat{\varphi}'$ are the coordinates of a supernova as seen by an off-center observer in the sky. Solving  (\ref{rs2}) for $a$, and substituting the outcome back to (\ref{dist}), we thus express $\hat{r}'$ in terms of the redshift $z$ instead of the scale factor $a$. The result of this calculation substituted into (\ref{dl}) for model IIA gives the luminosity distance expressed in terms the redshift $z$, the parameters of the model $\Omega_{inh}$, $w$, $r_0$, $\theta_0$, $\varphi_0$, $H_0$ and the angles $\hat{\theta}'$, and $\hat{\varphi}'$ at which a supernova is seen by an observer:
\be
\label{dl2}
d_L=\frac{(1+z)}{1-\frac{a_0 H_0^2 \Omega_{inh}}{4}r_0^2}~\hat{r}'(\Omega_{inh},w,r_0,\theta_0,\varphi_0, H_0,\hat{\theta}',\hat{\varphi}',z)~.
\ee
We can realize that in the limit $\Omega_{inh} \to 0$, the formula (\ref{dl2}) reduces to the standard flat Friedmann model filled with a single matter component which satisfies a barotropic equation of state.

\subsection{Model IIB}

In this subsection we consider another model of type II \citep{dabrowski93,dabrowski98} which is basically the same as the \citet{WdL} model. It has a different type of inhomogeneity than model IIA. We start with the same metric (\ref{Steph1}), but instead of assuming that the barotropic equation of state is fulfilled at the center of symmetry, we take exact forms of the scale factor and the curvature function as
\be
a(t) = \sigma t^{2/3},\ k(t)=-\alpha \sigma a(t),\ ~,
\ee
where the units of the constants are: $[\alpha]=(s/km)^{2/3}Mpc^{-4/3}$, $[\sigma]=(km/s)^{2/3}Mpc^{1/3}$, and time is measured in inverse to Hubble parameter units $[t]=s Mpc/km$.

The Einstein equations for such a model are given by \citep{dabrowski93}
\bea
\label{roWdL}
\frac{8\pi G}{c^4} \rho(t) &=& \frac{4}{3t^2} - \frac{3\alpha}{t^{2/3}}~, \\
\label{pWdL}
\frac{8\pi G}{c^4} p(t,r) &=& \frac{2\alpha}{t^{2/3}} - \frac{4}{3} \frac{\alpha \sigma^2}{t^{4/3}} r^2 + \alpha^2 \sigma^2 r^2~,
\eea
from which we can immediately see that at the center of symmetry $r=0$ no barotropic equation of state is fulfilled. An analytic form of the equation of state at the center of symmetry is instead
\be
\label{EOS1}
\rho = p \left( \frac{32 \pi^2 G^2}{3\alpha^3 c^8} p^2  - \frac{3}{2} \right)~~.
\ee
The equation (\ref{EOS1}) can also be written down as
\be
\rho + \frac{3}{2} p = \frac{c^4}{6\pi G t^2}~~.
\ee
The model approaches the dust-filled Friedmann universe if $\alpha \to 0$. The equation of state (\ref{EOS1}) may be fitted to the ideal gas interpretation of the inhomogeneous pressure \citep{sussman00}. From (\ref{pWdL}) one can see that there is a finite density singularity of pressure at $r \to \infty$.

We now follow the same procedure as for the previous model IIA. Applying the definition of conformal time (\ref{conftime})
\be
d \tau=\frac{d t}{a}=\frac{1}{\sigma}t^{-2/3} dt~,
\ee
and the condition (\ref{geod}) we have
\be
\tau=\hat{r}'(a) = \frac{3}{\sigma}\left(t_o^{1/3}-t_e^{1/3}\right)=\frac{3}{\sigma^{3/2}}\left(a_0 - a_e^{1/2}\right),
\ee
where $t_e$ and $t_o$ are the times of emission and observation, respectively.
The luminosity distance for the model IIB then reads as
\be
d_L=\frac{\sigma t_0^{2/3} (1+z)}{1 - \frac{1}{4} \alpha \sigma^2 t_0^{2/3}r_0^2} \hat{r}'(a)~~,
\ee
where the redshift (\ref{rs}) is
\be
1+z = \frac{\sigma t_0^{2/3}}{a_e} \frac{1 - \frac{1}{4} \alpha \sigma^2 t_e^{2/3}r_e^2}{1 - \frac{1}{4} \alpha \sigma^2 t_0^{2/3}r_0^2}
\ee
with $r_e$ given by (\ref{re}).

\section{Constraining the position of an observer with supernovae data}
\label{data}

We used a Bayesian framework based upon the Metropolis-Hastings MCMC (Markov Chain Monte Carlo) method to constrain the position of an off-center observer in the Stephani models IIA and IIB with the supernovae (SNIa) data. We took the likelihood function to be Gaussian in the form
\be
p({\rm data} | \Theta) \propto \exp ( - \frac{1}{2} \chi^2),
\ee
where $\Theta$ denotes the parameters of the considered models and ``data" denotes the SNIa data.  For the SNIa data $\chi^2$ takes the form
\be
\chi^2_{\rm SN}=\sum^{N}_{i,j=1} \left( C^{-1} \right)_{ij} \left[(\mu_{\rm obs}(z_i,\hat{\theta}'_i,\hat{\varphi}'_i)-\mu_{\rm pred}(z_i,\hat{\theta}'_i,\hat{\varphi}'_i))^2 \right]
\left[(\mu_{\rm obs}(z_j,\hat{\theta}'_j,\hat{\varphi}'_j)-\mu_{\rm pred}(z_j,\hat{\theta}'_j,\hat{\varphi}'_j))^2 \right]~.
\ee

\begin{figure}
\begin{center}
      \includegraphics[width=10.3cm]{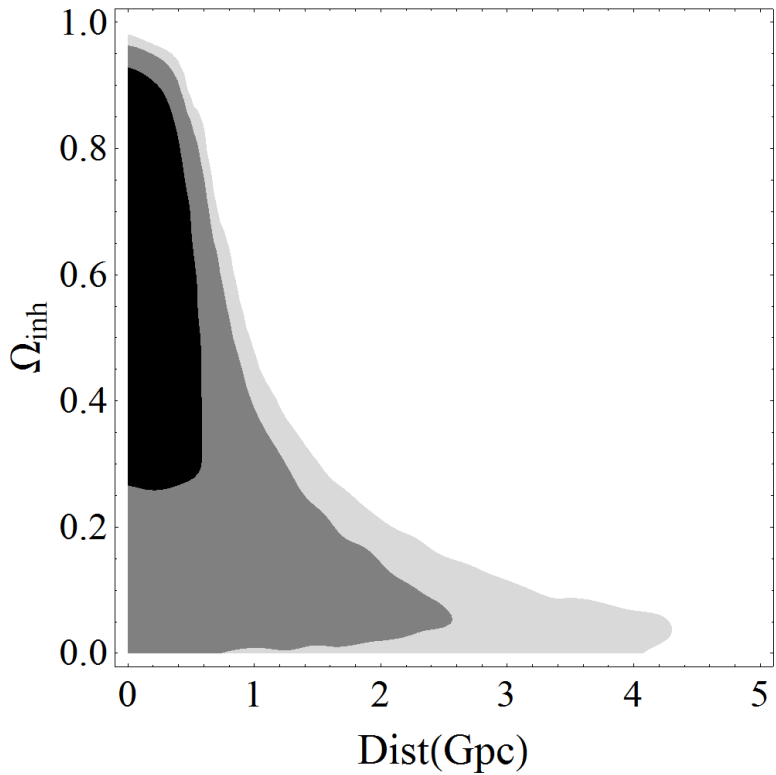}

\caption{The marginalized confidence intervals for inhomogeneous pressure model IIA in the dimensionless inhomogeneity density $\Omega_{inh}$ versus proper distance of an off-center observer position $Dist$ plane. The contours denote roughly 68\%, 95\% and 99\%  credible regions. It is clear that the position of an observer cannot be larger than about 4.4 Gpc away from the center and that it is smaller for larger inhomogeneity $\Omega_{inh}$. The most likely position of an off-center observer is $Dist=$ 341 Mpc $(\chi^2 = 526)$. Note that the homogeneous limit $\Omega_{inh} \to 0$ is effectively possible under the shift of the distance to the center being at infinity. This however, is equivalent to having strongly negative pressure matter to fill in the universe which plays the role of dark energy.}
\label{figuraOD}
\end{center}
\end{figure}

\begin{figure}
\begin{center}
\includegraphics[width=10.3cm]{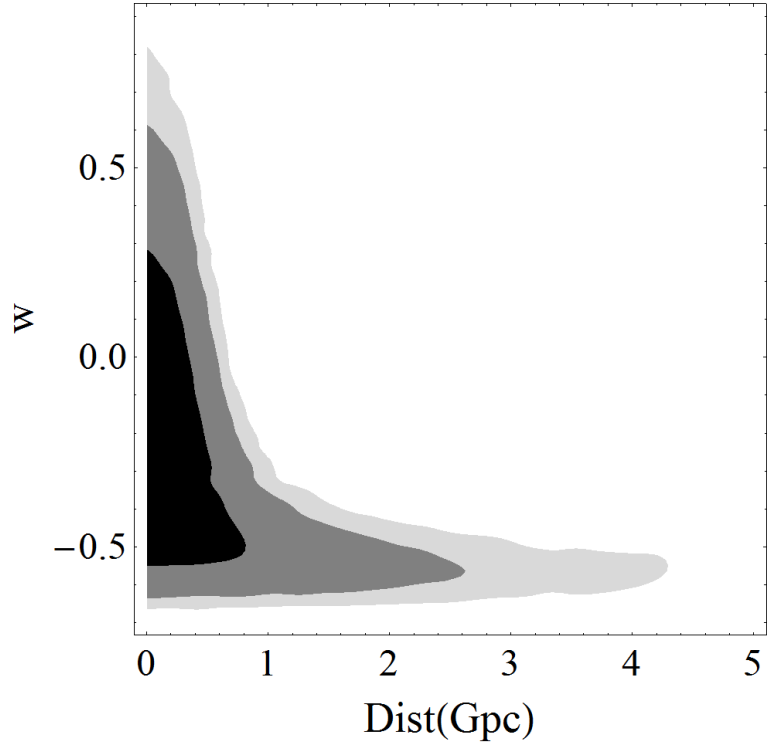}

\caption{The marginalized confidence intervals for inhomogeneous pressure model IIA in the center of symmetry barotropic index $w$ versus proper distance of an off-center observer position $Dist$ plane. The contours denote roughly 68\%, 95\% and 99\%  credible regions. One sees that more distant position of an observer is connected with having more and more negative pressure matter at the center of symmetry and that larger inhomogeneity prevents the observer from being too far from the center.}
\label{figurawD}
\end{center}
\end{figure}

\begin{figure}
\begin{center}

\includegraphics[width=10.3cm]{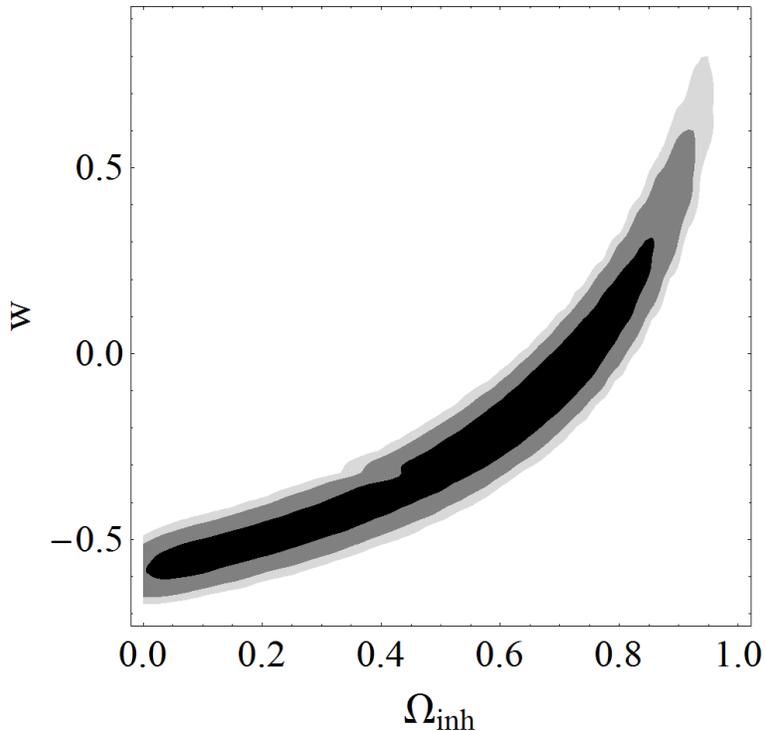}

\caption{The marginalized confidence intervals for inhomogeneous pressure model IIA in the center of symmetry barotropic index $w$ versus the dimensionless inhomogeneity density $\Omega_{inh}$. The contours denote roughly 68\%, 95\% and 99\%  credible regions. It is obvious that larger inhomogeneity is accompanied with higher positive pressure matter being allowed at the center of symmetry and so the inhomogeneity mimics the acceleration of the universe. On the contrary, more negative pressure matter at the center of symmetry is accompanied to a small inhomogeneity and this matter is the driving force for cosmic acceleration. The inhomogeneity density most likely value is $\Omega_{inh} = 0.77$ and the equation of state of the matter at the center of symmetry most likely value is $w = 0.093$ $(\chi^2 = 526)$.}
\label{figurawO}
\end{center}
\end{figure}

In Figs. \ref{figuraOD}-\ref{figurawO} we plot the contours which show the most likely position of an off-center observer in Gigaparsecs in inhomogeneous pressure Stephani universe IIA as limited by the Union2 sample of N=557 supernovae \citep{Union2}. The position is measured in terms of a proper distance
\be
\label{proper}
Dist \equiv \int_0^{r_0} \frac{a}{V} dr~~.
\ee
It is clear that at $1\sigma$ CL an observer cannot be further than at the distance of about $450 Mpc$, at $2\sigma$ CL he cannot be further than about 2.5 Gpc, and at $3\sigma$ further than about 4.4 Gpc in model IIA. It is apparently an approximate size of a void reported also for LTB models \citep{void1,void2}. From the plots, we can also conclude that the inhomogeneity density is non-zero and its most likely value is $\Omega_{inh} = 0.77$ (compare \citep{stephatests}). As for the equation of state of the matter at the center of symmetry, the value $w = 0.093$ is most favorable. The most likely position of an observer away from the center is $Dist=$ 341 Mpc $(\chi^2 = 526)$.

It also seems that more distant position of an observer is connected with having more and more negative pressure matter at the center of symmetry and that larger inhomogeneity prevents the observer from being too far from the center (see Fig. \ref{figurawD}). Larger inhomogeneity is accompanied by the higher positive pressure matter being allowed at the center of symmetry, and so the inhomogeneity mimics the current acceleration of the universe. On the contrary, more negative pressure matter at the center of symmetry is accompanied by small inhomogeneity, and such matter is the driving force for cosmic acceleration (see Fig. \ref{figurawO}).

\begin{figure}
\begin{center}
\begin{tabular}{cc}
       \\
\includegraphics[width=7.3cm]{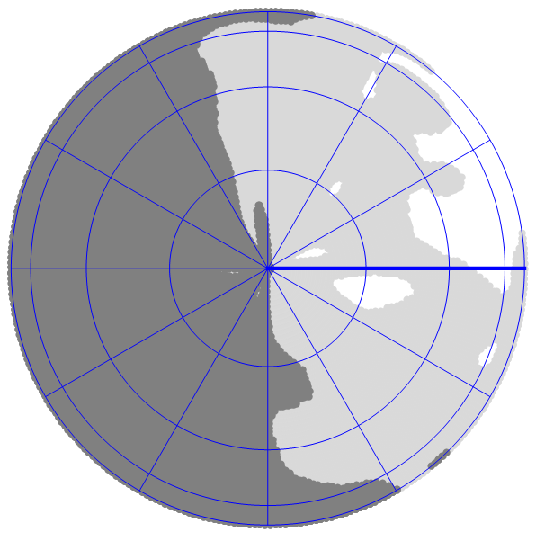} & \includegraphics[width=7.3cm]{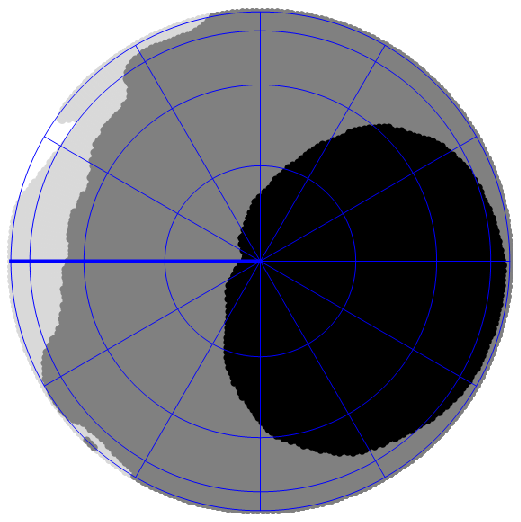}
\\
\end{tabular}

\caption{The position of the center of inhomogeneity for model IIA. On the left there is the North Celestial Hemisphere
and on the right is the South Celestial Hemisphere. The bold line is for zeroth Right
Ascension (meridian line). The most likely value of the Declination is $\delta = - 65.75^{\circ}$ and the Right Ascension is $a = 187.33^{\circ}$.}
\label{figurakaty}
\end{center}
\end{figure}

In Fig. \ref{figurakaty} we have plotted the confidence contours for the location of the center of inhomogeneity (center of symmetry) in the sky for the model IIA. On the left there is the North Celestial Hemisphere and on the right is the South Celestial Hemisphere. The bold line is for zeroth Right Ascension (meridian line). The center of inhomogeneity is placed at Declination $\delta = - 65.75^{\circ}$ and Right Ascension $a = 187.33^{\circ}$ and the distance to it from the observer is Dist = 341 Mpc.

\begin{figure}
    \begin{center}

      \includegraphics[width=10.3cm]{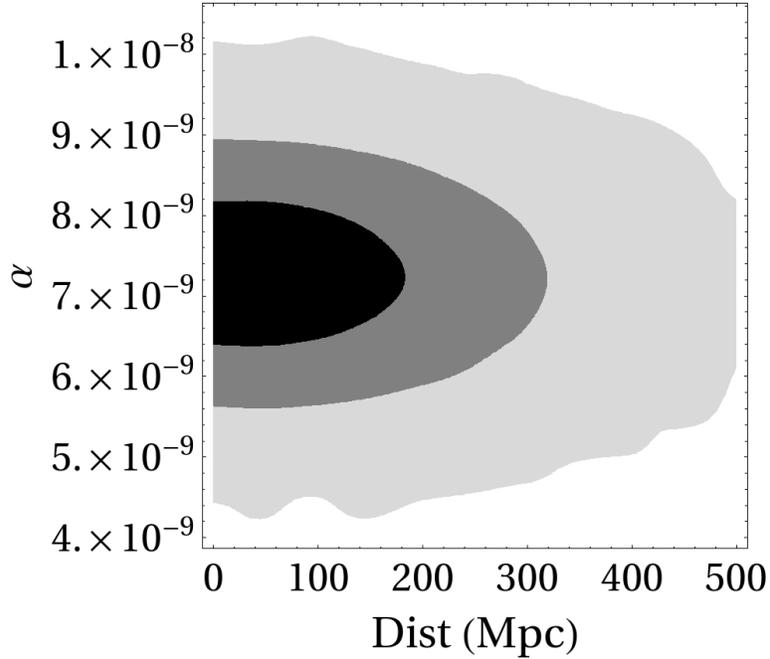}
\caption{The marginalized confidence intervals for inhomogeneous pressure model IIB in the inhomogeneity parameter $\alpha$ versus proper distance of an off-center observer position $Dist$ plane. The contours denote roughly 68\%, 95\% and 99\%  credible regions. The best-fit value of inhomogeneity parameter is $\alpha = 7.31 \cdot 10^{-9}$ $(s/km)^{2/3} Mpc^{-4/3}$. Note that this plot excludes the value of $\alpha \to 0$ since this is the dust limit (Einstein-de-Sitter) of the inhomogeneous model under study which is incompatible with supernovae data. The most likely value of the distance to the center is 68 Mpc ($\chi^2 = 557$).}
\label{figuraT}
\end{center}
\end{figure}

More restrictive results related to the position of an observer from the center are obtained for the model IIB. The off-center observer cannot be farther away from the center than about 215 Mpc at $1\sigma$ CL, 320 Mpc at $3\sigma$ CL, and 577 Mpc at $3\sigma$ CL. From the plot one sees that the inhomogeneity parameter is centered on an inhomogeneity parameter value of $\alpha = 7.31 \cdot 10^{-9}$ $(s/km)^{2/3} Mpc^{-4/3}$ $(\chi^2 = 557)$ which corresponds to the distance of 68 Mpc between the center of symmetry and an observer (see Fig. \ref{figuraT}).

\begin{figure}
\begin{center}
\includegraphics[width=7.3cm]{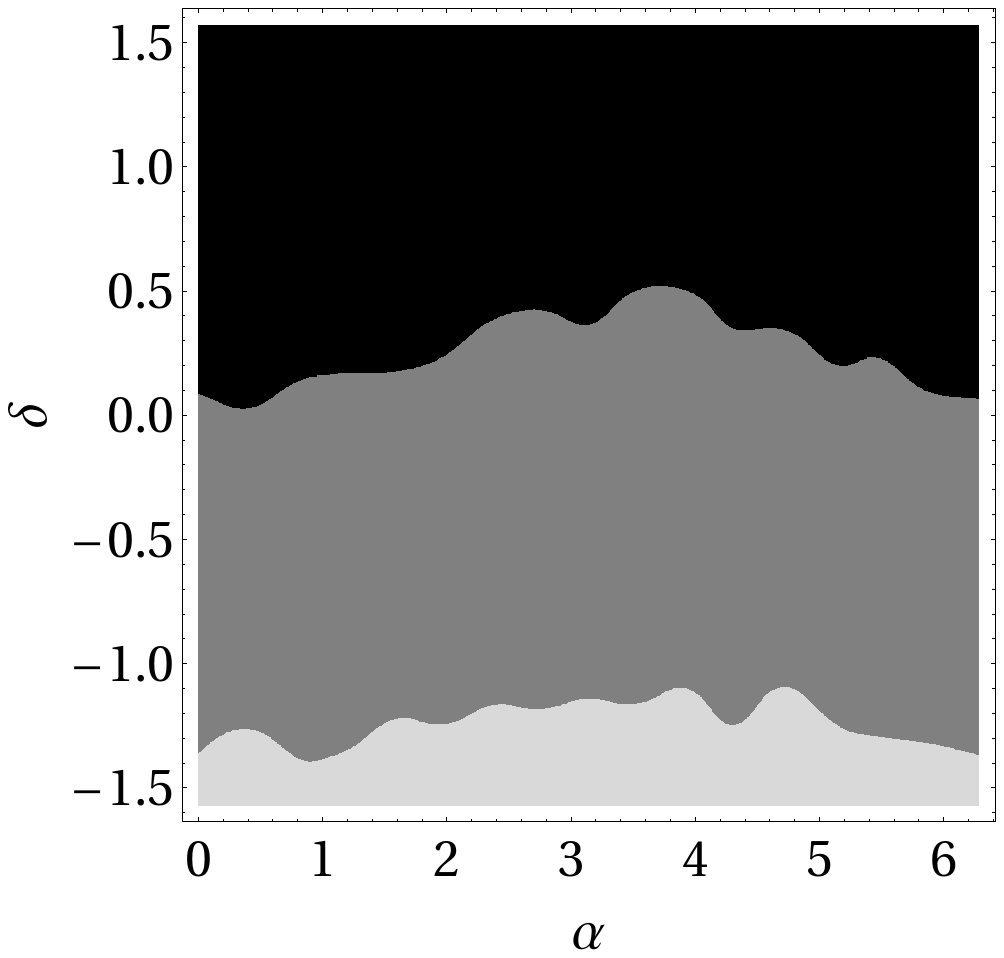}

\caption{The position (in radians) of the center of inhomogeneity for model IIB with respect to an Earth observer. The most likely values of the Declination is $\delta = 69.35^{\circ}$ and the Right Ascension is $a =8.39^{\circ}$.}
\label{figuraphitheta}
\end{center}
\end{figure}

The confidence contours for the location of the center of inhomogeneity for the model IIB are plotted in Fig. \ref{figuraphitheta}. With respect to an Earth observer the center is placed at Declination $\delta = 69.35^{\circ}$ and Right Ascension $a = 8.39^{\circ}$ and the distance to it is 68 Mpc.

\section{Results and Conclusions}
\label{conclusions}

We have presented exact formulas for the luminosity distance and the apparent magnitude of an astronomical object in inhomogeneous pressure Stephani universes for an off-center observer. Two specific Stephani models have been investigated. The first (marked as IIA) allowed for a barotropic equation of state to be valid at the center of symmetry with no exact function for the scale factor being specified. The second (marked as IIB) had no barotropic (though still analytic) form of an equation of state at the center, but its scale factor evolution was assumed to be exact and the same as for the dust-filled Friedmann universe. These models then represented different types of inhomogeneity - the fact which made our investigations more general.

Our exact luminosity distance and apparent magnitude formulas have then been applied to a sample data of Union2 supernovae \citep{Union2} in order to constrain possible position of an observer outside of the center of symmetry in these inhomogeneous pressure models.

Our results have shown that in model IIA an observer at $1\sigma$ CL cannot be further than about 450 Mpc away from the center, at $2\sigma$ CL he cannot be further than about 2.5 Gpc away, and at $3\sigma$ further than about 4.4 Gpc, which is comparable with evaluation of very large voids in LTB models \citep{chris2011,grande}. We have also found  that the inhomogeneity density has the most likely value $\Omega_{inh} = 0.77$ and the equation of state of the matter at the center of symmetry is characterized by a barotropic index value about $w = 0.093$. The most likely position of an observer away from the center is $Dist=$ 341 Mpc $(\chi^2 = 526)$.

More restrictive results related to the position of an observer away from the center have been obtained for the model IIB. The off-center observer cannot be farther away from the center than about 215 Mpc at $1\sigma$ CL, 320 Mpc at $3\sigma$ CL, and 577 Mpc at $3\sigma$ CL. We have also shown that the best-fit value of the inhomogeneity parameter is $\alpha = 7.31 \cdot 10^{-9}$ $(s/km)^{2/3} Mpc^{-4/3}$ which corresponds to the distance to the center of 68 Mpc $(\chi^2 = 557)$.

Model IIA has $\chi^2 = 526$ which is of 5 less than flat $\Lambda$CDM ($\chi^2 = 530.7$). Then, it is consistent with the data according to both Akaike and Bayesian information criteria, but not preferred over $\Lambda$CDM since to be so $\chi^2$ should be lowered by 8 (4 new parameters times 2) in the former case and by 25 (4 new parameters times $\ln{557}$) in the latter case. On the other hand, model IIB has $\chi^2$ of 26.3 greater than $\Lambda$CDM and so it is disfavored at more than $4 \sigma$.

We have also evaluated possible directions in the sky from the Earth to the center of inhomogeneity. For the model IIA it is at Declination is $\delta = - 65.75^{\circ}$ and the Right Ascension is $a = 187.33^{\circ}$ while for the model IIB it is Declination is $\delta = 1.21$ rad $= 69.35^{\circ}$ and the Right Ascension is $a = 0.15$ rad $=8.39^{\circ}$.

Though we do not take into account the local motions of an observer (who is comoving) with respect to the CMB in our models it might be interesting to ask if such directions may coincide with the directions of the Local Group (LG) motion claimed to appear at the velocity $V_{lg} = 627 \pm 22$ $km s^{-1}$ toward $(l,b) = (276^{\circ} \pm 3^{\circ}, 30^{\circ} \pm 3^{\circ})$ in galactic coordinates \citep{kogut,nusser} being more recently the matter of investigations of low-redshift local supernovae by \citet{feindt}. Our results in galactic coordinates give $(l,b) = (300.66^{\circ},-2.98^{\circ})$ for model IIA and $(l,b) = (121.35^{\circ},6.53^{\circ})$ for model IIB which does not seem to be very conclusive as far as possible alignment is concerned.

We have to emphasize that our tests have been based on supernovae data only. We have not discussed any other cosmological tests such as the CMB shift parameter, baryon acoustic oscillations, and the Sandage-Loeb redshift drift. Usually, supernovae do not impose such strong constraints onto the models as the CMB tests, so we think that the restrictions for the position of an off-center observer may even be more severe once taking them into account.

\acknowledgements

We would like to express our gratitude to Marek Kowalski and Uli Feindt for consulting the list of Union 2 supernovae with directions in the sky. MPD acknowledges the discussions with Roberto Sussman. The project was financed by the National Science Center Grant DEC-2012/06/A/ST2/00395. We thank the Referee for helpful suggestions. 

\appendix

\section{Non-isotropic versus isotropic radial coordinates}

The inhomogeneous Stephani metric (\ref{STMET}) uses the so-called isotropic coordinate $\bar{r}$ which is analogous to the isotropic coordinate applied in homogeneous and isotropic Friedmann-Robertson-Walker metric (and can be obtained from (\ref{STMET}) in the limit $k(t) \to k_0 = 0, \pm 1$) as follows
\be
\label{FRWbr}
ds_{\bar{r}}^2 = - c^2 dt^2 + \frac{a^2(t)}{V^2(\bar{r})}  \left[ d\bar{r}^2 + \bar{r}^2 \left(d \theta^2 + \sin^2{\theta} d\phi^2 \right) \right]~~,
\ee
where
\be
V(\bar{r}) = 1 + \frac{1}{4}k_0 \bar{r}^2~~,
\ee
in contrast to the most intensively used non-isotropic coordinate $r$, i.e.,
\be
\label{FRWr}
ds_{r}^2 = - c^2 dt^2 + a^2(t) \left[ \frac{dr^2}{1-k_0r^2}+ r^2 \left(d \theta^2 + \sin^2{\theta} d\phi^2 \right) \right]~~,
\ee
The relations between these coordinates are \citep{narlikar} (note that in the above formulas we have interchanged the meaning of coordinate $r$ from the Stephani metric (\ref{STMET}) into $\bar{r}$ in order to adopt standard notation which commonly uses $r$ for the non-isotropic coordinate)
\bea
r &=& \frac{\bar{r}}{1 + \frac{1}{4}k_0\bar{r}^2} = \frac{\bar{r}}{V(\bar{r})}~~,\\
\bar{r} &=& \frac{2r}{1 + \sqrt{1-k_0r^2}}~~.
\eea
Note that for $k_0=+1$, $\bar{r} \in (0,2)$, $r \in (0,1)$; for $k_0=0$, $\bar{r} \in (0,\infty)$, $r \in (0,\infty)$; for $k_0=-1$, $\bar{r} \in (0,2)$, $r \in (0,\infty)$. Usually, one defines a unified for the three curvature radial coordinate $\chi$ as follows
\bea
\label{rschi}
r = \frac{\bar{r}}{V(\bar{r})} = S(\chi) &=& \left\{
\begin{array}{l}
\label{Skchi}
\sin {\chi },\hspace{0.5cm}k_0=+1, \\
\chi, \hspace{0.5cm} k_0=0,  \\
\sinh {\chi },\hspace{0.5cm}k_0=-1,
\end{array}
\right. \
\eea
and so
\bea
\label{drschi}
\frac{dr}{d\chi} &=& \frac{dS(\chi)}{d\chi} = \left\{
\begin{array}{l}
\label{drSkchi}
\cos {\chi }\hspace{0.5cm}k_0=+1, \\
1 \hspace{0.5cm} k_0 =0,  \\
\cosh {\chi }\hspace{0.5cm}k_0=-1,
\end{array}
\right. \
\\
&=& \sqrt{1-k_0r^2} = \sqrt{1-k_0 S^2(\chi)}
\\
&=& \left\{
\begin{array}{l}
\label{drSkchi1}
\sqrt{1-\sin^2{\chi}} = \sqrt{1-r^2} \hspace{0.5cm}k_0=+1, \\
1 \hspace{0.5cm} k_0 =0,  \\
\sqrt{1+\sinh^2{\chi}} = \sqrt{1+r^2} \hspace{0.5cm}k_0=-1,
\end{array}
\right. \
\eea
On the other hand, we have
\be
\frac{dS(\chi)}{d\bar{r}} = \frac{dS(\chi)}{d\chi} \frac{d\chi}{d\bar{r}} = \frac{1 - \frac{1}{4}k_0\bar{r}^2}{\left( 1 + \frac{1}{4}k_0\bar{r}^2 \right)^2}~~,
\ee
and we can invert it as
\be
\frac{d\chi}{d\bar{r}} = \frac{dS(\chi)}{d\bar{r}} \frac{1}{\frac{dS(\chi)}{d\chi}}~~,
\ee
where
\be
\frac{dS(\chi)}{d\chi} = \sqrt{1-k_0 S^2(\chi)} = \frac{1 - \frac{1}{4}k_0\bar{r}^2}{1 + \frac{1}{4}k_0\bar{r}^2}~~,
\ee
and so
\be
\frac{d\chi}{d\bar{r}} = \left(1 + \frac{1}{4}k_0\bar{r}^2 \right)^{-1} = \frac{1}{V(\bar{r})}~~,
\ee
which means that
\be
\label{rtochi}
\frac{d\bar{r}}{V(\bar{r})} = d\chi~~.
\ee
It is useful to have the derivatives of one coordinate with respect to the other as follows
\be
\frac{d\bar{r}}{dr} = \frac{1}{\frac{dr}{d\bar{r}}} = \frac{1}{\frac{dS(\chi)}{d\bar{r}}} = \frac{\left( 1 + \frac{1}{4}k_0\bar{r}^2 \right)^2}{1 - \frac{1}{4}k_0\bar{r}^2}~~,
\ee
The application of the coordinate transformation given by (\ref{rtochi}) allows to transform the Stephani metric (\ref{STMET}) to the form analogous to that of non-isotropic coordinate Friedmann metric (\ref{FRWr}) (compare \citep{sussman00}), i.e.,
\bea
\label{STMETchi}
ds^2~=~-~\frac{a^2}{\dot{a}^2} \left[ \frac{ \left( \frac{V_{\chi}}{a} \right)^{\centerdot}}
  { \left( \frac{V_{\chi}}{a} \right)} \right]^2
c^2 dt^2~
+ \frac{a^2}{V_{\chi}^2} \left[d\chi^2~+~S^2(\chi)d\Omega^2
 \right],\nonumber \\
 &&
\eea
where
\be
\label{Vchi}
  V_{\chi} =  1 + k(t)S^2(\chi/2)~,
\ee
and
\bea
\label{S2chi}
S(\chi/2) &=& \left\{
\begin{array}{l}
\label{S2kchi}
\sin {\frac{\chi}{2} },\hspace{0.5cm}k_0=+1, \\
0, \hspace{0.5cm} k_0=0,  \\
\sinh {\frac{\chi}{2} },\hspace{0.5cm}k_0=-1,
\end{array}
\right. \
\eea
Using the non-isotropic coordinate (\ref{rschi}) one may express (\ref{S2chi}) as
\be
S(\chi/2) = \frac{1}{\sqrt{2}}\left( 1 - \sqrt{1-k_0 r^2} \right)^{1/2}~~,
\ee
and so the metric (\ref{STMET}) can be expressed in the non-isotropic coordinate as follows
\bea
\label{STMETr}
ds^2~=~-~\frac{a^2}{\dot{a}^2} \left[ \frac{ \left( \frac{V_{r}}{a} \right)^{\centerdot}}
  { \left( \frac{V_{r}}{a} \right)} \right]^2
c^2 dt^2~
+ \frac{a^2}{V_{r}^2} \left[\frac{dr^2}{1 - k_0r^2}~+~r^2 d\Omega^2
 \right],\nonumber \\
 &&
\eea
where
\be
\label{Vr}
  V_{r} =  1 + \frac{k(t)}{2} \left(1 - \sqrt{1-k_0 r^2} \right)~.
\ee

Note that for models II $((k/a)^{\centerdot} = 0)$ we obtain a simpler metric, which is an analogue of the metric (\ref{Steph1}).

\section{Null tangent vectors, conformal transformations and the radial isotropic coordinate}

It is possible to transform the non-isotropic Friedmann-Roberstson-Walker (FRW) coordinate metric (\ref{FRWr}) into the flat Minkowski metric by using the conformal transformation
of the form
\be
ds_r^2 = \Phi^2 ds_M^2~~,
\ee
with $\Phi$ being the conformal factor and where
\be
ds_M^2 = - c^2 dT^2 + dR^2 + R^2 \left(d \theta^2 + \sin^2{\theta} d\phi^2 \right)~~.
\ee
In terms of the $\chi$ coordinate the Friedmann metric (\ref{FRWr}) reads as
\be
ds_{\chi}^2 = - c^2 dt^2 + a^2(t) \left[ d\chi^2 + S^2(\chi) \left( d\theta^2 + \sin^2{\theta} d\phi^2 \right) \right]~~,
\ee
which by a simple coordinate transformation of the form
\be
c dt = c dT = a(\tau) d\tau
\ee
can be presented as
\be
\label{chimet}
ds_{\chi}^2 = a^2(\tau) d\bar{s}_{\chi}^2 = a^2(\tau) \left[ - d\tau^2 + d\chi^2 + S^2(\chi) \left( d\theta^2 + \sin^2{\theta} d\phi^2 \right) \right]  ~~,
\ee
and so $\Phi(\tau) = a(\tau)$ is the conformal factor.

The tangent vector to a null geodesic components in Minkowski space are solved easily as
\bea
&& k^{T}_M = \frac{dT}{ds} = 1, \hspace{0.3cm} k^R_M = \frac{dR}{ds} = \pm \sqrt{1 - \frac{h^2}{R^2}}, \nonumber \\
&& k^{\theta}_M = \frac{d\theta}{ds} = 0, \hspace{0.3cm} k^{\phi}_M = \frac{d\phi}{ds} = \frac{h}{R^2}~~.
\eea
The tangent vectors for the metric $d\bar{s}^2$ in (\ref{chimet}) are given by
\bea
&& k^{\tau} = \frac{d\tau}{ds} = 1, \hspace{0.3cm} k^{\chi} = \frac{d\chi}{ds} = \pm \sqrt{1 - \frac{h^2}{S^2(\chi)}}, \nonumber \\
&& k^{\theta}_M = \frac{d\theta}{ds} = 0, \hspace{0.3cm} k^{\phi}_M = \frac{d\phi}{ds} = \frac{h}{S^2{\chi}^2}~~.
\eea
The transformation rule for the tangent vectors reads as \citep{hawk_ellis}
\be
k^{\mu}_{FRW} = \Phi^{-2} \frac{\partial x^{\mu}}{\partial \tilde{x}^{\nu}} k^{\nu}_M~~,
\ee
where $\partial x^{\mu}/\partial \tilde{x}^{\nu}$ includes a coordinate transformation from coordinates $x^{\mu}$ to $\tilde{x}^{\nu}$ necessary to bring the metric
into a flat form so that we have
\bea
k^t_{FRW} &=&  \Phi^{-2} \frac{dT}{d\tau} \frac{d\tau}{ds} = \frac{1}{a}~,\\
k^r_{FRW} &=&  \Phi^{-2} \frac{dr}{d\chi}\frac{d\chi}{ds} = \pm \frac{\sqrt{1-k_0r^2}}{a^2} \sqrt{1 - \frac{h^2}{r^2}}~~,\\
k^{\theta}_{FRW} &=& \Phi^{-2} k^{\theta}_M = 0~~,\\
k^{\phi}_{FRW} &=& \Phi^{-2} k^{\phi}_M = \frac{h}{a^2r^2}~~.
\eea
In terms of the isotropic $\bar{r}$ coordinate we have
\bea
k^{\bar{r}}_{FRW} &=& k^r_{FRW} \frac{d\bar{r}}{dr} = \pm \frac{V}{a^2} \sqrt{1 - h^2 \frac{V^2}{\bar{r}^2}}~~,\\
k^{\phi}_{FRW} &=& h \frac{V^2}{a^2\bar{r}^2}~~.
\eea

\end{document}